# Finetuning Lightweight LLMs for Control Flow Graph Generation

Hanyu Zhang
Dept. of Industrial and Management Systems Engineering
Waseda University
Tokyo, Japan
bankzhy@akane.waseda.jp

Tomoji Kishi
Dept. of Industrial and Management Systems Engineering
Waseda University
Tokyo, Japan
kishi@waseda.jp

*Abstract*— **Control Flow Graph (CFG) is an important program representations for software analysis, code understanding, and software maintenance. Traditional CFG generation techniques mainly rely on bytecode or abstract syntax trees. However, these approaches usually require complete, compilable, and syntax-error-free code, which limits their applicability to incomplete or erroneous code. Furthermore, they often depend on language-specific tools, making it difficult to support multiple programming languages in a unified manner. To address these limitations, this paper investigates the use of fine-tuned lightweight large language models (LLMs) for CFG generation. We first design a unified CFG output format and a task-specific fine-tuning prompt for CFG generation. Then, we construct a dataset based on an existing LeetCode dataset through automatic CFG generation and error augmentation. We evaluate the proposed approach on six lightweight LLM models, including three code-specific LLMs: CodeLlama, QwenCoder, and DeepSeekCoder; and three general-purpose LLMs: Llama3.2-3B, Qwen-4B, and Phi-4B. The experimental results show that, through fine-tuning, lightweight LLMs achieve promising results for CFG generation, particularly when the input code is incomplete or erroneous. It also demonstrates cross-language generalization capability on programming language not included in the fine-tuning data.**

## I. Introduction

A Control Flow Graph (CFG) is a graph-based representation of all possible execution paths in a program, where nodes denote basic statements and edges represent the transfer of control between them. Since its theory was pointed out by Allen [1], CFG has become one of the most fundamental intermediate representations in program analysis and software engineering (SE). It serves as the structural foundation for a wide range of SE downstream tasks [2], including static analysis, compiler optimization, 32

Over the past decades, numerous approaches and tools have been proposed to automatically generate CFGs from source code. Existing approaches could generally be categorized into two main groups: bytecode-based approaches and Abstract Syntax Tree (AST)-based approaches. Bytecode-based approaches generate CFGs from compiled intermediate representations, leveraging precise low-level execution semantics. The representative example are Soot [4] and WALA [3], which generate CFGs from Java bytecode and both have been widely adopted in static program analysis. In contrast, AST-based approaches operate directly on source code syntax structures, enabling higher-level structural analysis before compilation. Representative examples include Spoon [5] for Java and Py2Cfg [24] for Python, both of which construct CFGs from parsed syntax trees. These tools have significantly advanced CFG construction by improving automation and supporting multiple programming languages.

However, existing CFG generation approaches still suffer from several practical limitations. For bytecode-based approaches, source code must be complete and compilable before the CFG construction. This strict requirement greatly limits their usefulness in realistic development scenarios, where code may be incomplete, partially written, or under active modification. AST-based approaches relax this requirement by operating on uncompiled code, but they still depend on successful syntax parsing. In the presence of syntax errors, malformed statements, or incomplete code fragments, AST construction may fail, directly causing the CFG generation failure or producing structurally incorrect CFGs. Moreover, even when code is syntactically valid, careless coding practices or accidental violations of coding conventions may introduce implicit semantic inconsistencies, causing the generated CFG to deviate from the developer's original intent. Recent work has highlighted that both explicit syntax errors and implicit semantic errors can significantly affect CFG correctness [6].

Furthermore, traditional CFG generation tools are usually language-specific, requiring different frameworks for different programming languages. For some languages, mature CFG generation tools may not even exist. These limitations motivate the need for a more flexible and language-agnostic approach for CFG generation directly from original source code.

Large Language Models (LLMs) have recently emerged as a powerful paradigm for understanding and generating natural language and source code. By training on massive-scale corpora, LLMs can capture both syntactic patterns and semantic relationships, enabling them to perform complex reasoning over code. In software engineering, LLMs have also been successfully applied to several software engineering tasks such as code completion, bug detection, program repair, code summarization, and code translation [8]. Unlike traditional parser-based approaches, LLMs exhibit strong tolerance toward grammatical irregularities, incomplete statements, and misspelled tokens, allowing them to process imperfect code inputs more robustly. More importantly, LLMs can infer contextual semantics and identify implicit logic inconsistencies, potentially reducing semantic deviations in related software engineering tasks. Specialized code-oriented LLMs such as OpenAI Codex [19], Claude Code [25] have demonstrated remarkable performance in programming tasks. However, these large-scale models often incur high training, inference, and deployment costs, making them less practical for resource-constrained environments. To address this issue, lightweight LLMs such as Code-Llama [20], Qwen-Coder [21], and DeepSeek-Coder [22] have also been proposed in recent years, offering a better balance between capability and efficiency.

Despite their potential, directly applying LLMs to CFG generation introduces several challenges. First, the output format of CFGs is often inconsistent and highly dependent on prompt design, making it difficult to standardize results across models and experiments. Second, different prompt formulations may produce structurally different graph representations for the same code. Some CFG representations may require a large number of output tokens, increasing inference cost and reducing efficiency. Moreover, the quality of generated CFGs can be highly unstable, with some outputs containing missing nodes, incorrect edges, or logically inconsistent control-flow structures. This instability becomes even more severe when using lightweight LLMs, whose reasoning and structural consistency are often weaker than larger models. Therefore, it requires further research on LLM-based CFG generation. Further discussion of this problem has also been provided in previous studies [6] [9] [10] [11].

In this study, we investigate the feasibility of fine-tuning lightweight LLMs for automatic CFG generation. To address the limitations of existing approaches, we first design a unified CFG representation that reduces token consumption while preserving essential control-flow semantics. Based on this representation, we further develop a unified prompting strategy to standardize CFG generation behavior across different models. Next, we construct a dataset based on an existing LeetCode dataset [26] through automatic CFG generation and error augmentation. Finally, we fine-tune six representative lightweight LLMs and conduct a comprehensive comparative evaluation to assess their CFG generation capability. The experimental results demonstrate the effectiveness of lightweight LLMs for CFG generation including the tolerance to incomplete or erroneous code and the cross-language generalization capability.

## II. Related Works

In this section, we will review some representative approaches for CFG generation, including traditional static analysis techniques and recent LLM-based approach.

The bytecode-based approach is a tradition CFG generation approach. A representative example is the work by Vallée-Rai et al. [4], which introduced *Soot*, a widely used Java analysis and optimization framework. Soot constructs CFGs by first compiling source code into an intermediate bytecode representation (e.g., Jimple), and then performing control flow analysis on this normalized form. The bytecode-based approach does effectively generate CFG, however its have strict prerequisites: the input code must be complete and compliable. As a result, any syntax errors or incomplete code fragments will prevent successful compilation, thereby causing CFG generation to fail and limiting the usefulness of such approaches in practical scenarios.

To overcome these limitations, AST-based approaches such as Spoon [5] or the recent work by Tran and Hung [7], have also been proposed. These approaches operate directly on source code by parsing it into an AST and then analyzing structural elements (e.g., statements and control constructs) to generate CFGs. Compared to bytecode-based approaches, These approaches relax the requirement of compilation and allows CFG generation at the source level. This makes it more flexible and suitable for static analysis tasks during early development stages. However, these approaches still rely on successful parsing of the source code. In the presence of syntax errors or incomplete code, AST construction may fail or produce incorrect structures, which in turn leads to inaccurate or failed CFG generation.

With the rapid advancement of LLMs, recent studies have also explored CFG generation using LLM. The representative work is the work proposed by Huang et al [6]. This approach leverages a chain-of-thought-style pipeline built on top of large LLMs (e.g., Codex [19]) to iteratively infer CFG structures from partial or incomplete code. The method demonstrates strong capability in handling incomplete inputs and reduces dependence on traditional parsing techniques. However, it relies on large proprietary models, resulting in high computational and deployment costs, which limits its practical usability.

In summary, traditional bytecode-based approaches impose strict requirements on code completeness and compellability, significantly limiting their applicability. AST-based approaches alleviate some of these constraints by operating directly on source code, but they still depend on syntactically correct inputs. Furthermore, existing tools are often language-specific, and the availability of robust CFG generation tools varies across programming languages. With the emergence of LLMs, researchers have begun exploring CFG generation through prompt-based approach. However, these approaches typically rely on general-purpose large code models and prompt engineering, which incur substantial deployment costs.

In this paper, we propose a novel CFG generation approach based on fine-tuning lightweight LLMs. The proposed approach removes strict constraints on input code, enabling CFG generation even for incomplete or erroneous code. By focusing on lightweight models, our approach significantly reduces computational requirements and deployment costs, making it more practical for real-world applications.

## III. Approach

### A. Overview

Our approach aims to generate control flow graphs from source code by fine-tuning lightweight large language models. Given a source code snippet as input, the objective is to produce a structurally correct CFG that represents the execution flow among statements. In our approach, first, we define a unified CFG representation to eliminate the format inconsistency among different CFG generation tools. Second, we design a structured prompt to guide LLMs to generate CFGs in the required format. Third, we construct a CFG generation dataset based on the existing *greengerong/leetcode* dataset [26], covering four programming languages (Java, Python, C++ and JavaScript). During dataset construction, we automatically generate CFG labels and further introduce error-augmented code samples to simulate realistic programming scenarios. Finally, we fine-tune six lightweight LLMs, including three code-specific LLMs and three general-purpose LLMs, to evaluate the effectiveness and generalizability of the proposed method. The detail is explained in following subsections.

## B. Unified CFG Representation

The CFG could be represented not only as graphical structures but also in various textual formats, such as JSON or XML. However, existing CFG generation tools, such as Soot [4], Spoon [5], and Py2CFG [24], do not adopt a unified output format, and they often differ in how they represent control-flow paths for conditional and loop structures. Such format inconsistency makes it difficult to directly compare CFGs generated by different tools or use them as unified training labels for LLM-based generation.

In addition, since LLM-based generation is sensitive to output length, the target CFG representation should be compact and token-efficient. Therefore, we adopt a simplified digraph-style representation inspired by Graphviz [23]. This representation explicitly describes nodes and edges while avoiding redundant metadata. The general schema of the proposed representation is shown in Figure.1.

```
digraph <method_name> {{
    ...
    <id> [label="<statement>", shape="<shape>"]
    ...
    <from> -> <to>
    ...
}}
```

Figure 1. Unified CFG representation schema

In this format, *method_name* denotes the name of the target method or function. Each *id* uniquely identifies a CFG node according to the original statement in the source code. The *label* field records the source-code statement represented by the node, while the *shape* field indicates the type of the statement. Specifically, *rectangle* is used for ordinary process statements, *diamond* for conditional statements, *hexagon* for loop statements, and *parallelogram* for return or output statements. Finally, *<from> -> <to>* defines a directed edge, representing the possible transfer of control from one node to another. To illustrate the proposed CFG representation, Figure.2 presents an example Python implementation of the classic *twoSum* algorithm. Figure.3 shows the corresponding graphical CFG representation, and Figure.4 presents its textual digraph representation

```
def twoSum(nums, target):
    map = {}
    for i, num in enumerate(nums):
        complement = target - num
        if complement in map:
            return [map[complement], i]
        map[num] = i
    return []
```

Figure 2. Example code of CFG representation

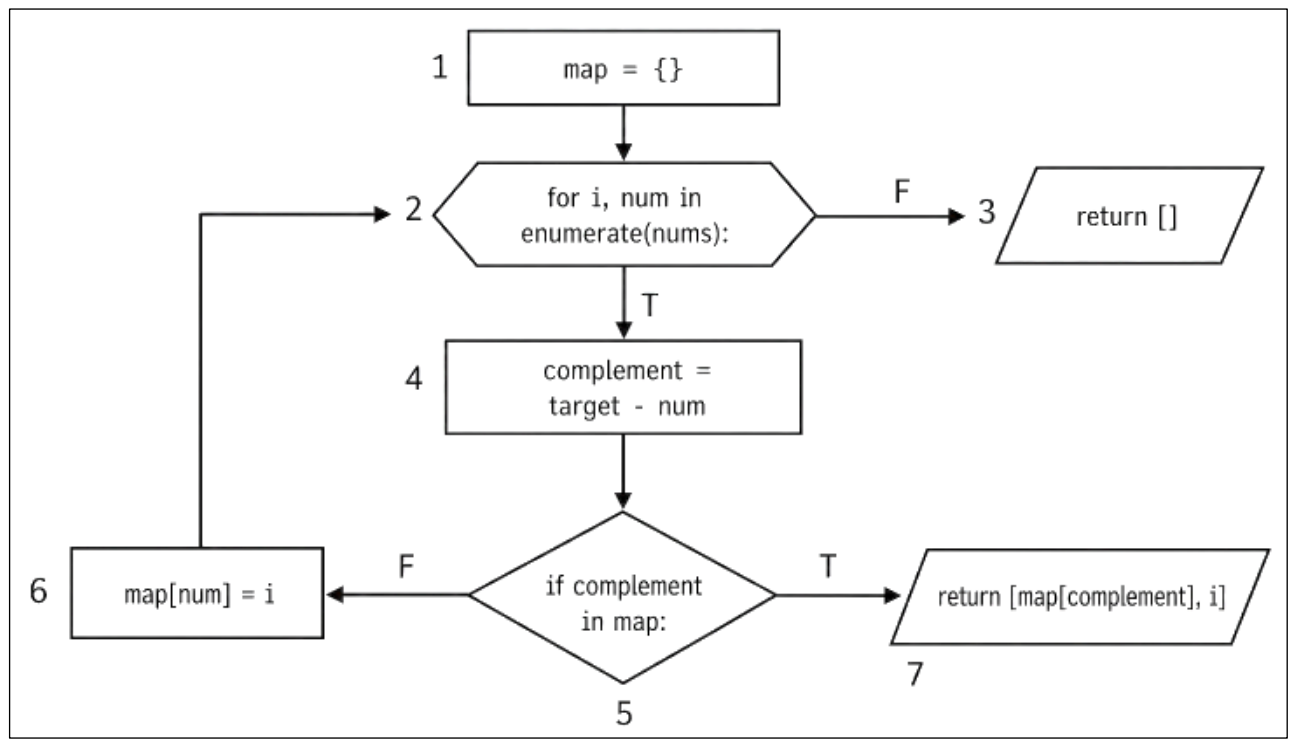

Figure 3. Control flow graph of the example code

```
digraph twoSum {
    1 [label="map = {}", shape="rectangle"]
    2 [label="for i, num in enumerate(nums):", shape="hexagon"]
    3 [label="return []", shape="parallelogram"]
    4 [label="complement = target - num", shape="rectangle"]
    5 [label="if complement in map:", shape="diamond"]
    6 [label="map[num] = i", shape="rectangle"]
    7 [label="return [map[complement], i]", shape="parallelogram"]

    1 -> 2
    2 -> 3
    2 -> 4
    4 -> 5
    5 -> 6
    5 -> 7
    6 -> 2
}
```

Figure 4. Digraph representation of the CFG

## C. Prompt Design

To improve the stability of LLM outputs and enhance their ability to generate structured CFGs, we design a task-specific prompt for CFG generation. The prompt consists of four parts: task instruction, CFG construction rules, output requirements, and an example. The task instruction explicitly defines the role and objective of the model. Specifically, the model is instructed as follows: "*You are a control flow graph generator. Task: Generate the Control Flow Graph (CFG) for the given method.*" This instruction helps the model focus on graph generation rather than code explanation or program summarization. The CFG construction rules define the basic elements of a CFG, including statement types and the connection rules between statements.

The output requirements are designed to constrain the generated content. The model is required to output only the CFG, without explanations, markdown formatting, or additional natural language text. This strict output constraint is important for automatic evaluation because any additional text may affect CFG parsing and metric calculation. Finally, an example is included in the prompt to demonstrate the expected input-output pattern. The example follows the unified CFG format described in previous subsections.

### D. Dataset Construction

We construct the experimental dataset based on the *greengerong/leetcode* [26] dataset, which contains a large number of programming solutions for algorithmic problems. The dataset includes implementations in four programming languages, including Java, Python, C++, and JavaScript. The reason why we use this base dataset is it covers diverse algorithmic logic, such as loops, conditionals, early returns, and nested control structures. And the size of this dataset is suitable for performing an initial evaluation.

Based on the original dataset, we perform two major processing steps. The first step is automatic CFG generation. For each code sample in original dataset, we generate the CFG for its three languages Java, JavaScript and Python. We use TreeSitter [27] to parse the source code into an AST. Then, following common AST-to-CFG construction logic according to the existing AST-based CFG generation approaches [5] [7], we traverse the AST and convert control structures into nodes and directed edges. The generated CFGs are further normalized into the unified digraph format defined in Section III-B.

The second step is code error augmentation. To evaluate whether the fine-tuned LLMs can handle imperfect code, we generate four erroneous variants for each original code sample. The injected errors include missing symbols, typographical errors, undefined variables, and incomplete statements. Importantly, each erroneous variant shares the same CFG label as its original clean sample because these minor errors are designed not to change the intended control-flow structure. This strategy simulates real-world development scenarios, where code fragments may be incomplete or contain small errors during editing. By introducing such samples into the training data, the model is encouraged to learn the intended control-flow semantics rather than relying only on fully parsable code.

### E. Model Fine-tuning

We select six lightweight LLMs as experimental subjects as two groups. The first group consists of code-specific lightweight LLMs, with parameter size around 7B, including CodeLlama-7B [20], Qwen2.5-Coder 7B [21], and DeepSeek-Coder 6.7B [22]. These models are designed or optimized for code-related tasks. The second group consists of general-purpose lightweight text-only LLMs, with parameter size around 4B, including Llama-3.2-3B-Instruct [29], Phi-4-mini-instruct [30], and Qwen3-4B [31]. Although these models are not specifically designed for code related-tasks, they provide a useful comparison for examining whether general lightweight LLMs can also acquire CFG generation capability through fine-tuning. All selected models are fine-tuned using supervised instruction tuning.

## IV. Evaluation Experiment

### A. Experimental Setup

The experimental dataset used in this study was constructed following the procedure described in Section III-D. Specifically, for each sample, we performed CFG generation and error augmentation on the corresponding code in three programming languages: Java, Python, and JavaScript. For dataset splitting, since the original dataset contains approximately 2.36K rows. In our experiment, approximately 2,000 original samples and their corresponding erroneous variants were used as the training set, while the remaining samples were used as the test set. Furthermore, to evaluate the cross-language CFG generation capability of the fine-tuned models, only Java and Python code samples were used for training, whereas JavaScript samples were excluded from training and used only for testing. As results, we got 13,331 training samples and 3,583 test samples.

We fine-tuned all selected lightweight LLMs using the Unsloth [28], which provides an efficient framework for parameter-efficient model fine-tuning. For each model, the number of training epochs was set to 3 to ensure a consistent experimental setting across all models. All experiments were conducted on a workstation with NVIDIA RTX 6000 GPU. The same prompt template, training strategy, and evaluation procedure were applied to all models to ensure fair comparison.

### B. Evaluation Metrics

To comprehensively evaluate the quality of the generated CFGs, we calculate metrics at two levels: the node level and the edge level. At each level, we will compute five metrics: Precision, Recall, F1-score, EMR (Exactly Matched Ratio), and ZMR (Zero Matched Ratio). Node-level metrics measure whether the generated CFG contains the correct executable statements, while edge-level metrics evaluate whether the control-flow among statements are correctly predicted.

Precision evaluates the correctness of generated nodes and edges, while Recall measures how many ground-truth elements are recovered. The F1-score balances both metrics and is therefore used as the primary measure of node-level and edge-level CFG generation accuracy. EMR measures the proportion of samples whose predicted CFG exactly matches the ground-truth CFG. At the node level, this means that all predicted nodes are completely consistent with the ground-truth nodes; at the edge level, it means that all predicted control-flow are exactly correct. In contrast, ZMR measures the proportion of samples for which no valid node or edge can be matched with the ground truth. A higher ZMR indicates that the generated CFG is largely unusable. Therefore, EMR and ZMR provide sample-level evidence of complete success and complete failure, respectively.

### C. Evaluation Questions

The experiments are designed to answer the following research questions. **Q1. What is the overall performance on the test dataset?** This question evaluates the general CFG generation capability of fine-tuned lightweight LLMs on the Java and Python test dataset, including both original and error-augmented samples. It aims to verify the effectiveness of fine-tuning and compare the overall performance of different models.

**Q2. Does erroneous code affect CFG generation performance?** This question examines whether source-code errors affect CFG generation by comparing performance on clean and erroneous samples in test dataset. The result reflects the robustness of fine-tuned models when handling imperfect or incomplete code. **Q3. What is the cross-language generalization ability on a non-finetuned programming language, JavaScript?** This question evaluates whether models fine-tuned on Java and Python can generalize to JavaScript. It helps examine whether the models learn transferable control-flow patterns rather than only language-specific features.

### *D. Evaluation Results*

**Q1. What is the overall performance on the test dataset?** After the fine-tuning process, Figure.5 shows the test results of each model on the test dataset, which includes both original and error-augmented samples in Java and Python. From the results, **QwenCoder achieved the best performance**, with the highest Node F1 (**0.95**) and Edge F1 (**0.65**). **Qwen-4B ranked second**, with a Node F1 of **0.93** and an Edge F1 of **0.64**. **DeepseekCoder performed the worst**, with the lowest Node F1 (**0.52**) and the highest Node-ZMR (**0.41**). A possible reason for the poor performance of DeepseekCoder is that it produced more zero-matched samples, which means many predictions had no matched CFG nodes with the reference results. This may be caused by unstable output format or failure to follow the required CFG generation format.

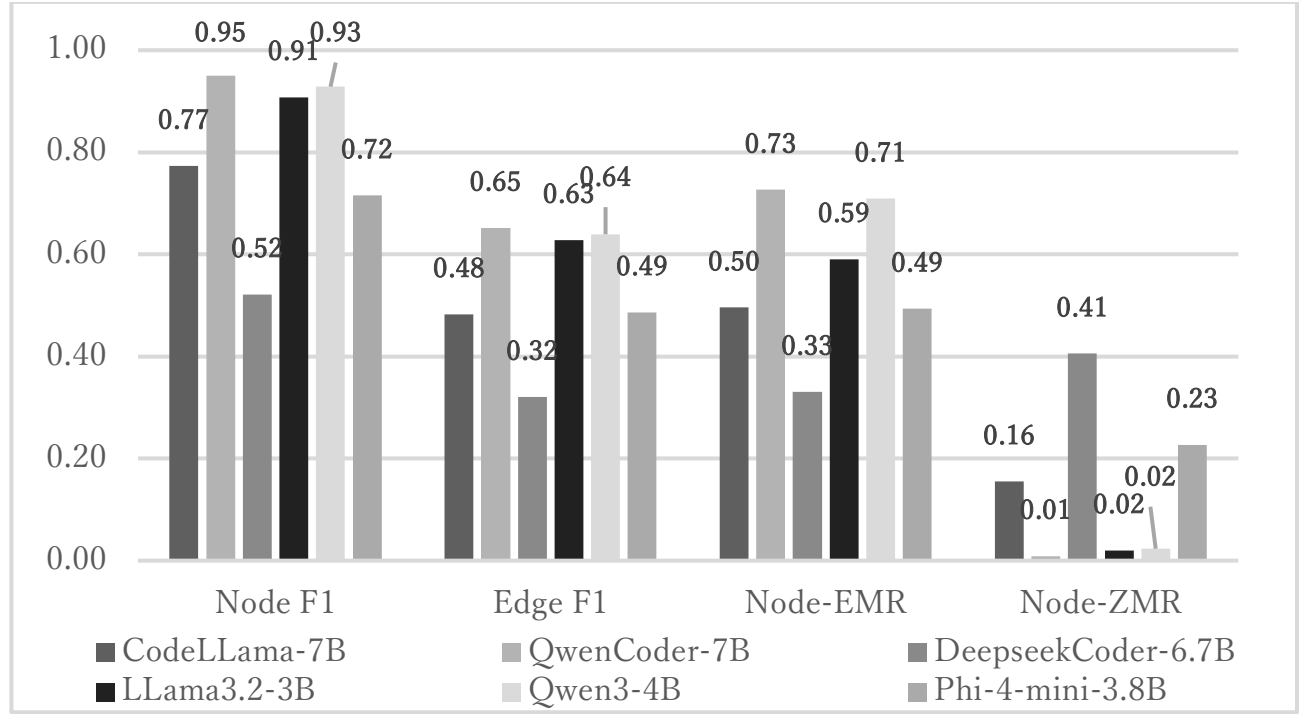

Figure 5. Overall performance on test dataset

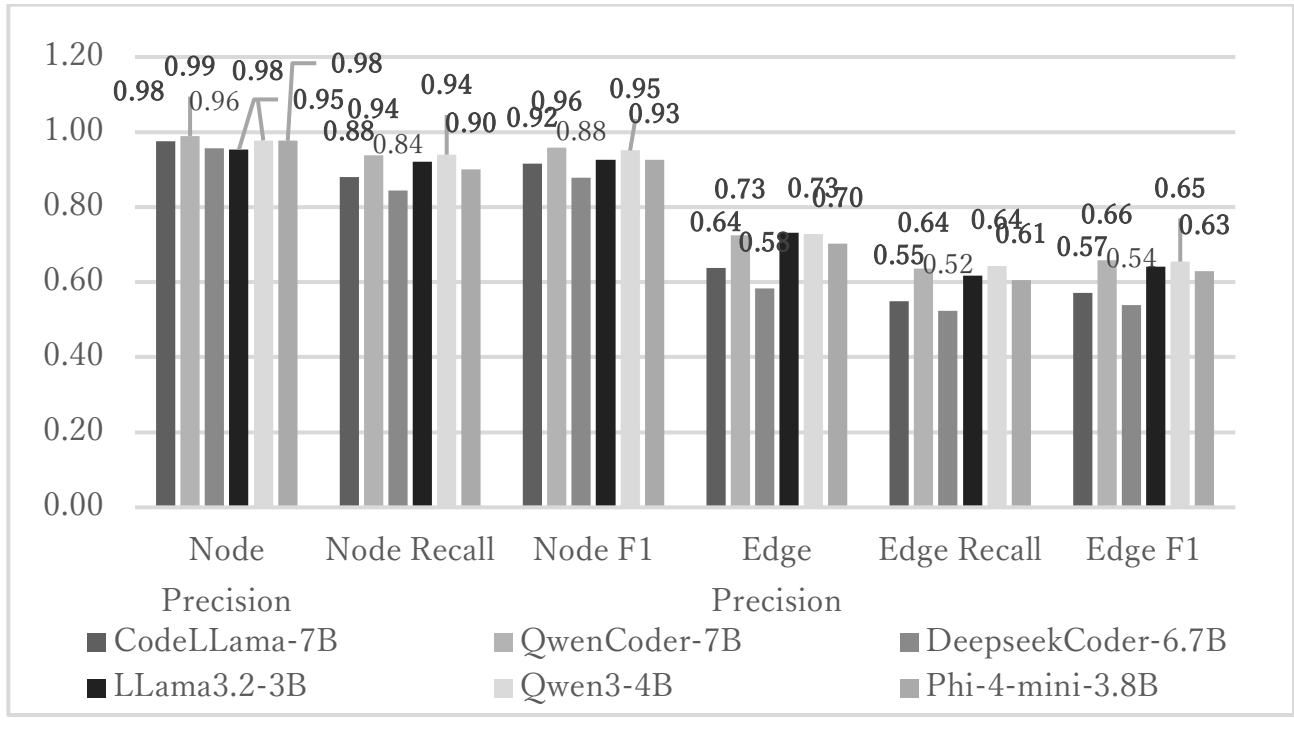

Figure 6. ZMR Samples Removed performance

Furthermore, we removed the zero-matched samples, where the predictions contained no valid matched nodes or edges, and recalculated the performance. As shown in Figure.6, **QwenCoder still achieved the best overall performance**, with a Node F1 of 0.96 and an Edge F1 of 0.66. Qwen-4B ranked second, with a Node F1 of 0.95 and an Edge F1 of 0.65. DeepSeekCoder remained the weakest model, with a Node F1 of 0.88 and an Edge F1 of 0.54. However, its performance improved substantially compared with the original results. This suggests that DeepSeekCoder can generate reasonable CFGs for some samples, but its overall performance is strongly affected by a large number of zero-match outputs.

**Q2. Does erroneous code affect CFG generation performance?** To answer this question, we compute the performance metrics for both clean and erroneous code samples and visualize their differences in Figures.7 and Figures.8, where the differences are calculated as clean-code metrics minus erroneous-code metrics. The results show that the differences are relatively small across most models, generally below 0.01, indicating that the evaluated models exhibit a certain degree of robustness to erroneous code.

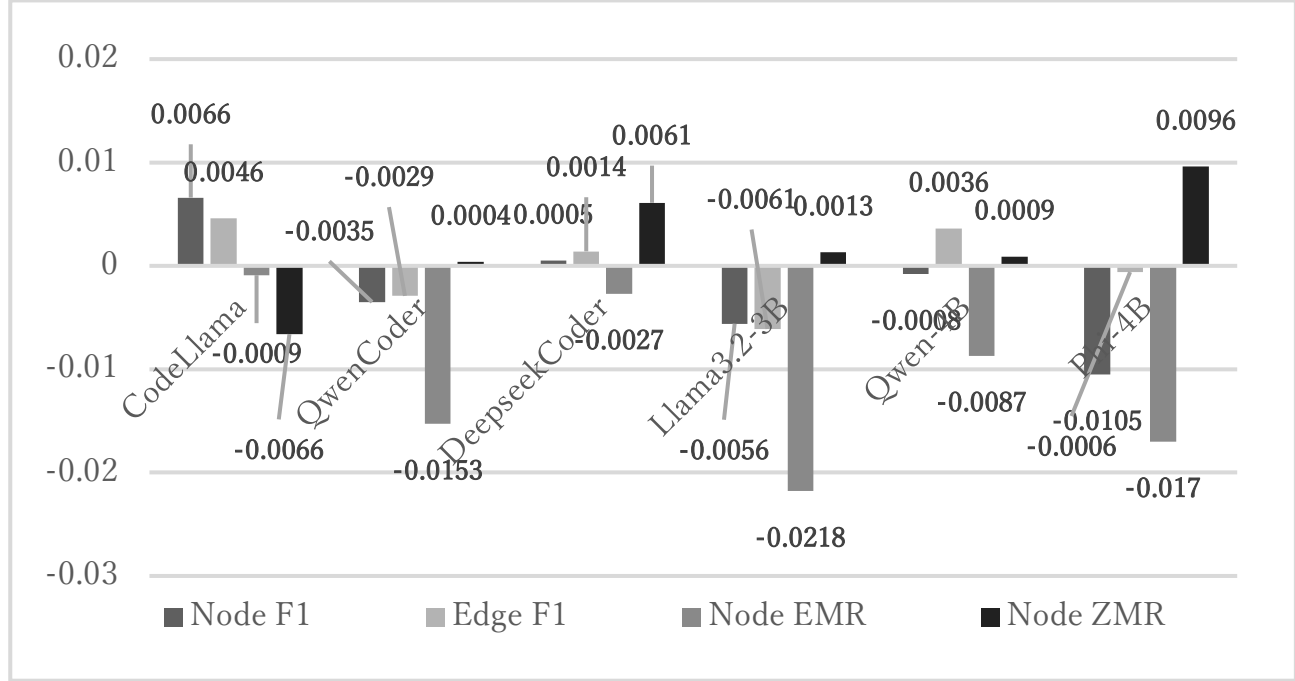

Figure 7. The Difference of overall performance

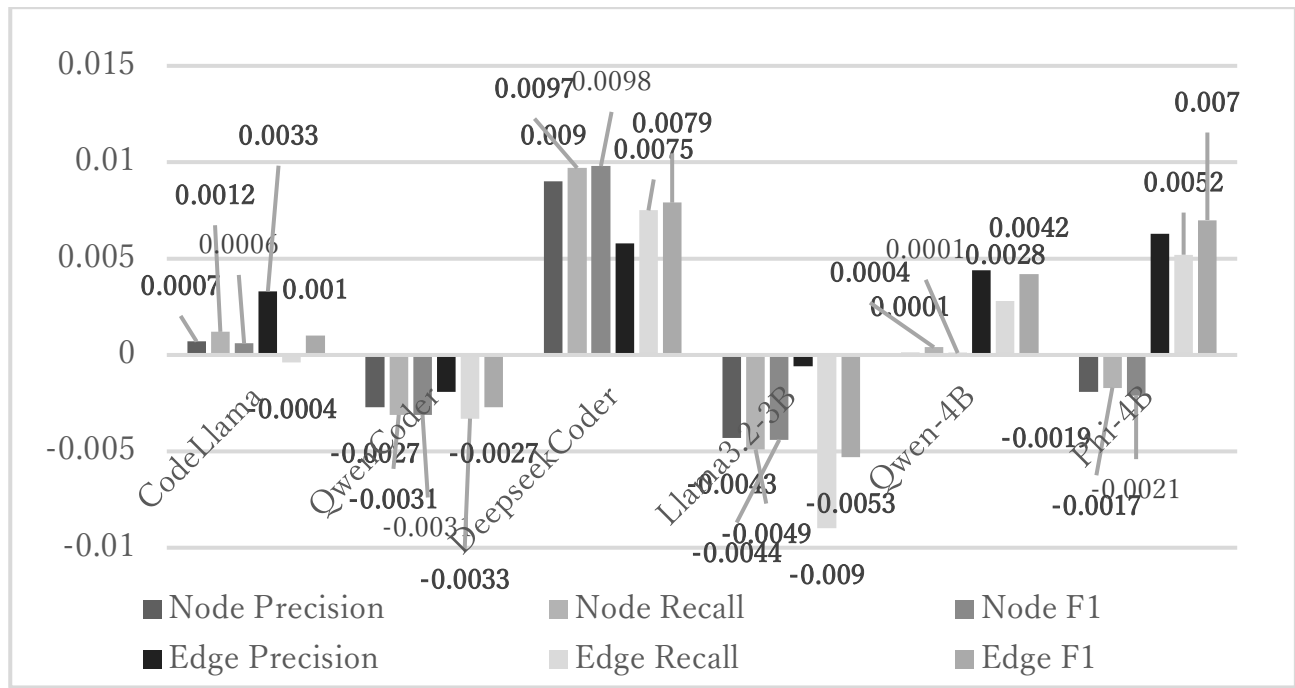

Figure 8. The Difference after ZMR-Samples Removed

**Q.3 What is the cross-language generalization ability on an non-finetuned programming language, JavaScript?** To answer this question, we evaluate the fine-tuned models on the JavaScript test dataset, which was not included in the fine-tuning languages. Figure.9 presents the results, where Node F1 and Edge F1 are calculated after removing zero-matched samples. Compared with the results in Figure.6, the performance on JavaScript decreases by approximately 0.20 overall, indicating that CFG generation on an unseen programming language remains more challenging.

Nevertheless, in terms of F1-score, all six models retain a certain degree of CFG generation capability on JavaScript. This suggests that the fine-tuned models can generalize to a non-finetuned programming language to some extent. However, the ZMR values reveal clear differences in output stability. In particular, DeepSeekCoder and CodeLlama show noticeably higher ZMR values than the other models, indicating that they are more likely to produce completely unmatched or unusable CFG outputs. Therefore, although their successful predictions can still achieve reasonable F1-scores, their overall reliability on JavaScript is limited by unstable generation behavior.

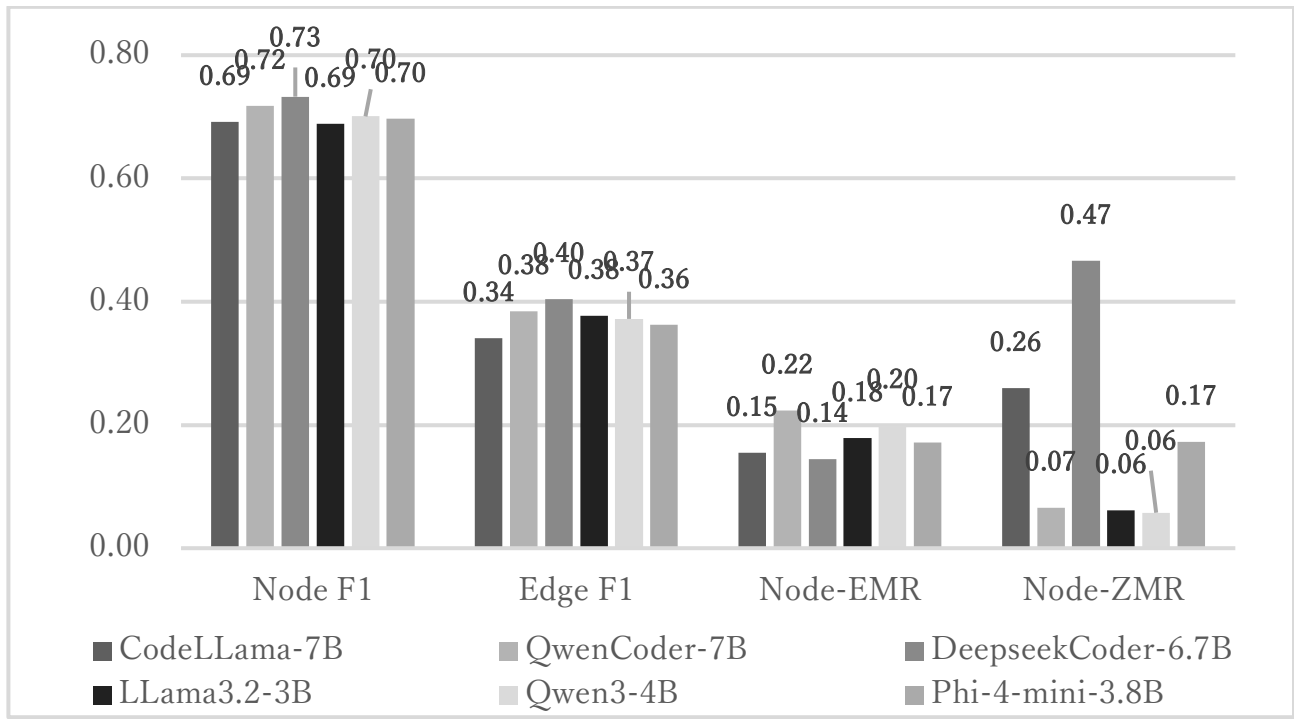

Figure 9. The overall performance on JavaScript dataset

## V. CONCLUSION

In this paper, we explored the use of fine-tuned lightweight large language models for CFG generation. We first designed a unified CFG representation and a task-specific prompting strategy to standardize CFG generation across different models. Next, we constructed a dataset based on an existing LeetCode dataset through automatic CFG generation and error augmentation. Finally, we fine-tuned six representative lightweight LLMs and conducted a comprehensive comparative evaluation to assess their CFG generation capability. The experimental results demonstrate that fine-tuned lightweight LLMs are effective for CFG generation, which could adapt to incomplete or erroneous code, and exhibit cross-language generalization capability on programming languages not included in the fine-tuning data. In future work, we will conduct larger-scale experiments to further evaluate the performance of the proposed approach in practical development scenarios.